\documentstyle[12pt]{article}
\begin{document}
\normalsize
\begin{center}
{\Large \bf $sl(2)$ VARIATIONAL SCHEMES FOR SOLVING ONE CLASS OF
NONLINEAR QUANTUM MODELS}\\

\vspace{7mm}
               V.P. KARASSIOV\\
{\it Lebedev Physical Institute, Leninsky prospect 53, 117924 Moscow,
Russia\\
E-mail: vkaras@sci.lpi.msk.su;  karas@sci.fian.msk.su}
\end{center}
\begin{abstract}
Hamiltonians of a wide-spread class of $G_{inv}$-invariant nonlinear quantum
models, including multiboson and frequency conversion ones, are expressed as
non-linear functions of $sl(2)$ generators. It enables us to use standard
variational schemes, based on  $sl(2)$ generalized coherent states as trial
functions, for solving both spectral and evolution tasks. In such a manner
a new analytical expression is found for energy spectra in a mean-field
approximation which is beyond quasi-equidistant ones obtained earlier.
\end{abstract}

\section{Introduction}

For last decades a great attention has been paid to developments of both
exact and approximate techniques to solve and examine different dynamical
problems for quantum strongly coupled systems whose interaction Hamiltonians
are expressed by nonlinear functions of operators describing subsystems
(see, e.g., [1-9] and references therein). However, as a rule, such
techniques either are adapted for treating special forms of model
Hamiltonians and initial quantum states [1-5,7-9] or require lengthy and
tedious calculations (as it is the case, e.g., for the algebraic Bethe
ansatz [6]).

Recently, a new universal Lie-algebraic approach has been developed [10-13]
to get exact solutions of both spectral and evolution problems for some 
nonlinear quantum models of strongly coupled subsystems having symmetry groups
$G_{inv}$. It was based on exploiting a formalism of polynomial Lie algebras 
$g_{pd}$ as dynamic symmetry algebras $g^{DS}$ of models under study, and, 
besides, generators of these algebras $g_{pd}$ can be interpreted as 
$G_{inv}$-invariant "essential" collective dynamic variables in whose terms 
model dynamics are described completely. Specifically, this approach enabled 
us to develop some efficient techniques for solving physical tasks in the 
case of $g^{DS}=sl_{pd}(2)$, when model Hamiltonians $H$ are expressed as 
follows
$$ H = aV_0 +g V_+ + g^* V_- +C,\quad [V_{\alpha}, C]=0,
\quad V_- =(V_+)^+,      \eqno (1.1) $$
where $C$ is a function of model integrals of motion $R_i$ and $V_0, V_{\pm}$
are the $sl_{pd}(2)$ generators satisfying the commutation relations
$$ [V_0, V_{\pm}]= \pm V_{\pm}, \quad [V_-, V_+] =  \psi_n(V_0+1) -
\psi_n(V_0),$$
$$\psi_n(V_0)=A\prod_{i=1}^{n} (V_0+\lambda_i(\{R_j\}))   \eqno (1.2)$$
The structure polynomials $\psi_n(V_0)$ depend additionally on $\{R_i, i=1,
\dots\}$, and their exact expressions for some wide-spread classes of
concrete models were given in [10-12].

All techniques [10-13] are based on using expansions of most important
physical quantities (evolution operators, generalized coherent states (GCS),
eigenfunctions etc.) by power series in the $sl_{pd}(2)$ shift generators
$V_{\pm}$ and on decompositions
$$ L(H) =\sum_{\oplus}L([l_i]), \quad (V_+V_- -\psi_n(V_0)\equiv -\psi_n(R_0)
|_{L([l_i])}=0        \eqno (1.3)$$
of Hilbert spaces $L(H)$ of quantum model states in direct sums of
the subspaces $L([l_i])$ which are irreducible with respect to joint actions
of algebras $sl_{pd}(2)$ and symmetry groups $G_{inv}$ and
describe specific "$sl_{pd}(2)$-domains" evolving independently in time
under  action of the Hamiltonians (1.1); $[l_0]$ are lowest weights
of $L([l_i]): \psi_n(l_0)=0$ and other quantum numbers $l_i, i=1, \dots$ are
eigennumbers of operators $R_i$. Then, using restrictions of Eqs. (1.1)-(1.2)
on $L([l_i])$, one can develop simple algebraic calculation schemes for
finding evolution operators
$$U_{H}(t)=\sum_{f=-\infty}^{\infty}V_+^f u(v_0;t), \quad V_+^{-f}\equiv
V_-^{f}([\psi_n(V_0)]^{(f)})^{-1}, \; [\psi_n(x)]^{(f)}\equiv
\prod_{r=0}^{f-1}\psi_n(x-r),   \eqno (1.4a)$$
amplitudes $Q_v(E_f)$ of expansions
$$ |E_f\rangle = A_f\prod_{j}(V_+ -\kappa^f_j) |[l_i]\rangle=
\sum_v  Q_v(E_f) |[l_i]; v\rangle    \eqno (1.4b)$$
of energy eigenstates $|E_f\rangle$ in
orthonormalized bases $\{|[l_i]; v\rangle : V_0 |[l_i]; v\rangle=
(l_0+v) |[l_i]; v\rangle\}$) and appropriate energy spectra $\{E_f\}$ of
bound states [11,13]. (In fact, the factorized form of $|E_f\rangle$ given
by the first equality in (1.4b) realizes
an efficient modification of the algebraic Bethe ansatz [6] in
terms of collective dynamic variables related to the $sl_{pd}(2)$
algebras [11,13].) In the paper [12] some explicit integral
expressions were found for amplitudes $Q_v(E)$, eigenenergies $\{E_a\}$
and "coefficients" $u(v_0;t)$ of evolution operators $U_{H}(t)$ with the
help of a specific "dressing" (mapping) of solutions of some auxiliary
exactly solvable tasks with the dynamic algebra $sl(2)$.

However, all exact results obtained do not yield simple working formulas
for analysis of models (1.1) and revealing different physical effects (e.g.,
a structure of collapses and revivals of the Rabi oscillations [2,8],
bifurcations of solutions [5] etc.) at
arbitrary initial quantum states of models. Therefore, it is necessary to
develop some simple techniques, in particular, to get some closed, perhaps,
approximate expressions for evolution operators, energy eigenvalues and wave
eigenfunctions, which would describe main important physical features of
model dynamics with a good accuracy (cf. [5,8,9]). Below we examine some
possibilities along these lines for models (1.1)-(1.2) by means
of reformulating them in terms of the formalism of the usual $sl(2)$ algebra
and developing variational schemes corresponding to quasiclassical
approximations for original models by analogy with developments [5,14-16].
 
\section {A reduction of linear $sl_{pd}(2)$ problems to non-linear
$sl(2)$ ones}

We can reformulate models (1.1)-(1.2) in terms of $sl(2)$ generators using an
isomorphism of the $sl_{pd}(2)$ algebras to extended enveloping algebras
${\cal U}_{\psi}(sl(2))$ of the familiar algebra $sl(2)$.
This isomorphism is established via a generalized Holstein-Primakoff mapping
given on each subspace $L([l_i])$ as follows [10,11]
$$Y_0 = V_0-l_0\mp j,\; Y_+= V_+ [\phi_{n-2}(Y_0)]^{-1/2},\;
\phi_{n-2}(Y_0)=\frac{\psi_n(Y_0+l_0\pm j+1)}{(j\mp Y_0)(\pm j+1+Y_0)},\;
Y_-=(Y_+)^+,$$
$$ [Y_0, Y_{\pm}]= \pm Y_{\pm}, \quad [Y_-, Y_+] = \mp2 Y_0 \eqno (2.1)$$
where $ Y_{\alpha}$ are the $sl(2)$ generators, $\mp j$ are lowest weights
of $sl(2)$ irreducible representations realized on subspaces
$L([l_i])$ and $\psi_2(x)=(j\pm x)(\pm j+1-x)$ are quadratic structure
functions $\psi_n(x)\equiv\psi_2(x)$ of $sl(2)$ (hereafter upper/lower signs
corresponding to the $su(2)$/$su(1,1)$ algebras are chosen for
finite/infinite dimensions $d([l_i])$ of the spaces $L([l_i])$).

Note that, by definition, functions $\phi_{n-2}(Y_0)$ on spaces $L([l_i])$
can be chosen as polynomials of $(n-2)$-th degree in $Y_0$. For example,
substituting
$$\psi_3 (V_0)=\frac{1}{4}(2V_0 +R_2-R_1)(2V_0 +R_1+R_2)(-V_0 +R_2+1),$$
$$l_0=\frac{|k|-s}{3}, \quad l_1=k,l_2=\frac{|k|+2s}{3}, \quad k=0,\pm1,
\pm2,...;\;s=2j= d([l_i])-1=0,1,2,... \eqno (2.2)$$
for three-boson models [11-13]
$$ H_{tb} = \omega_1 a^+_1 a_1 +\omega_2 a^+_2 a_2 +\omega_3 a^+_3 a_3 +
g(a^+_1a^+_2) a_3 + g^*(a_1a_2) a_3 ^+  ,    \eqno (2.3a)$$
$$V_0 =(N_1+N_2-N_3)/3,\; V_+ =(a^+_1a^+_2) a_3, \quad a= \omega_1 +\omega_2-
\omega_3, \quad N_i=a^+_i a_i, $$
$$2C =R_1(\omega_1 -\omega_2) +R_2(\omega_1+\omega_2 +2\omega_3),
\; R_1= N_1-N_2,\;3R_2= N_1+N_2+2N_3 \eqno (2.3b)$$
we get $\phi_{1}(Y_0)= Y_0+j+|k|+1$. Similar expressions can be found
for $\phi_{1}(Y_0)$ in the cases of the point-like Dicke  and the
second harmonic generation models taking appropriate expressions from [12].

Then, restrictions $H_{[l_i]}$ of Hamiltonians (1.1) on
$L([l_i])$ may be re-written in terms of $Y_{\alpha}$ as follows
$$ H_{[l_i]} = aY_0 + Y_+\tilde g(Y_0) + \tilde g^+(Y_0) Y_- +\tilde C,
\quad \tilde g(Y_0)= g\sqrt{\phi_{n-2}(Y_0)},\;\tilde C=C+a(\pm j+l_0),
   \eqno (2.4) $$
Evidently, this form corresponds to generalizations of semi-classical (linear
in $sl(2)$ generators) versions of matter-radiation interaction models
[8,9,12] by introducing operator (intensity-dependent) coupling coefficients
$\tilde g(Y_0)$ (cf. [3,7]). Emphasize, however, a collective (not
associated with a single subsystem) nature of operators
$Y_{\alpha}$ in Eq. (2.4) (cf. [9]); therefore, dynamic variables
$Y_{\alpha}$ correspond to a non-standard  quasiclassical
approximation (when $\tilde g(Y_0)=const$ in Eq. (2.4)) of original models
as it follows, e.g., from a direct comparison of such an approximation with
standard (when creation/destruction operators of one mode are replaced by
$c$-numbers) semiclassical limits for the model (2.3).

If $n=2$, then $\phi_{n-2}(Y_0)=1, sl_{pd}(2)=sl(2), l_0=\pm j $, and the
formalism of GCS related to the $SL(2)$ group displacement operators
$$S_Y(\xi=re^{i \theta})=\exp(\xi Y_+-\xi^* Y_-)=\exp[t(r)e^{i \theta} Y_+]
\exp[-2\ln c(r) Y_0]\exp[-t(r)e^{-i\theta} Y_-],
\eqno (2.5)$$
($t(r)= \tan r/\tanh r, c(r)=\cos r/\cosh r$ for $su(2)/su(1,1)$)
yields a powerful tool for solving both spectral and evolution tasks [16].

Specifically, in this case, using the well-known $sl(2)$ transformation
properties of operators $Y_{\alpha}$ under the action of $S_Y(\xi)$ [16]:
$$S_Y(\xi)Y_{+}S_Y(\xi)^{\dagger}\equiv Y_{+}(\xi)=[c(r)]^2 Y_{+}
\pm e^{-i\theta} [ s(2r) Y_0 - e^{-i\theta} [s(r)]^2 Y_-], \; Y_{-}(\xi)=
(Y_{+}(\xi))^{\dagger},$$
$$ S_Y(\xi)Y_{0}S_Y(\xi)^{\dagger}\equiv Y_{0}(\xi)=c(2r) Y_{0}
- \frac{s(2r)}{2} [ e^{i\theta} Y_{+} +  e^{-i\theta} Y_{-}, \quad  s(r) =
sin r/sinh r,         \eqno (2.6)$$
Hamiltonians $H_{[l_i]}$ can be transformed into the form
$$ \tilde{H}_{[l_i]}(\xi)=S_Y(\xi)H_{[l_i]}
S_Y(\xi)^{\dagger}=  \tilde C +Y_0 A_0(a, g; \xi)+Y_+ A_+(a, g; \xi) +
Y_- A^*_+(a, g; \xi)  \eqno (2.7a)$$
At the values $\xi_0=\frac{g}{2|g|}arctan\frac{2|g|}{a}$ for $su(2)$  and
$\xi_0= \frac{g}{2|g|}arctanh\frac{2|g|}{a}$ for $su(1,1)$ of the parameter
$\xi$ one gets $A_+(a, g; \xi)=0$, and the Hamiltonian
$\tilde{H}_{[l_i]}(\xi)$ takes the form
$$\tilde{H}_{[l_i]}(\xi_0)= \tilde C +Y_0\sqrt{a^2\pm 4 |g|^2}
\eqno (2.7b)$$
which is diagonal  on eigenfunctions $|[l_i];v\rangle=
\tilde N(j,v)(Y_+)^v|[l_i];v=0\rangle, \; N^{-2}(j,v)=v!(2j)!/(2j-v)!\;
\mbox{for}\; su(2)\;\mbox{and}\; N^{-2}(j,v)=v!\Gamma (2j+v)/\Gamma (2j)\;
\mbox{for}\; su(1,1)$. Therefore, original Hamiltonians $H_{[l_i]}$
have  the eigenenergies
$$E_{v}([l_i];\xi_0)= \tilde C +(\mp j+v)\sqrt{a^2\pm 4 |g|^2}
\eqno (2.8a)$$
and eigenfunctions
$$|[l_i];v;\xi_0\rangle=
S_Y(\xi_0)^{\dagger}|[l_i];v; \rangle \eqno (2.8b)$$

Similarly, when $sl_{pd}(2)=sl(2)$, operators $S_Y(\xi (t))$ are "principal"
parts in the evolution operators $U_{H}(t)= \exp(i\phi (t) Y_0) S_Y(\xi (t))$
with $c$-number functions $\phi (t), \xi (t)$ being determined from a set of
non-linear differential equations corresponding to classical motions [16,17].

However, for arbitrary degrees $n$ of $\psi_n(V_0)$ Hamiltonians (2.4)
are essentially non-linear in
$sl(2)$ generators $Y_{\alpha}$, and, therefore, the situation is very
changed.  Particularly, in general cases it is unlikely to diagonalize
$H_{[l_i]}$ with the help of operators
$S_Y(\xi)$ since analogs of Eq. (2.7a) on multi-dimensional spaces $L([l_i])$
$$\tilde{H}_{[l_i]}(\xi)=S_Y(\xi)H_{[l_i]}S_Y(\xi)^{\dagger}=
 aY_0 (\xi) + Y_+(\xi)\tilde g(Y_0(\xi)) + \tilde g^+(Y_0(\xi)) Y_- +\tilde C
    \eqno (2.9) $$
contain  (after expanding them in power series) many terms with higher
powers of $Y_{\pm}$ [13].

Nevertheless, the formalism of the $SL(2)$
group GCS $|[l_i];v;\xi\rangle=S_Y(\xi)^{\dagger}|[l_i];v; \rangle$ [16]
can be an efficient tool for analyzing non-linear models [5,11,14-16], in
particular, for getting approximate analytical solutions. Specifically, a
simplest example of such  approximations was obtained in [11] by mapping
(with the help of the change $V_{\alpha}\rightarrow Y_{\alpha}$)
Hamiltonians (1.1) by Hamiltonians $H_{sl(2)}$ which are linear in $sl(2)$
generators $Y_{\alpha}$ (but with modified constants $\tilde a, \tilde g$)
and have on each fixed subspace $L([l_i])$ equidistant energy spectra
obtained from Eq. (2.8a). However, this (quasi)equidistant approximation,
in fact, corresponding to a substitution of certain effective coupling
constants $\tilde g$ instead of true operator entities $\tilde g(Y_0)$ in
Eq. (2.4), does not enable to display many peculiarities of models (1.1)
related to essentially non-equidistant parts of their spectra. Therefore,
it is needed in corrections, e.g., with the help of iterative schemes
[8,14,15]; specifically, one may develop perturbative schemes by using
expansions of operator entities $\tilde g(Y_0)$ in Taylor series in
$Y_0$ as it was made implicitly for the Dicke model in [8,9]. But there
exist a more effective, incorporating many peculiarities of models (1.1),
way to amend the quasi-equidistant approximation.

\section{ $SL(2)$ energy functionals and variational schemes for solving
spectral and evolution tasks}

This way is in applying $SL(2)$ GCS $|[l_i];v;\xi\rangle=S_Y(\xi)^{\dagger}
|[l_i];v; \rangle$ as trial functions
in the variational schemes of determing energy spectra and quasiclassical
dynamics [5,15]. Indeed, the results (2.8) are  obtained by using a
variational scheme determined by the stationarity conditions
$$a)\;\frac{\partial {\cal H}([l_i];v;\xi)}{\partial \theta}=0,\quad
b)\;\frac{\partial {\cal H}([l_i];v;\xi)}{\partial r}=0
\eqno (3.1)$$
for the energy functional ${\cal H}([l_i];v;\xi)=\langle[l_i];v;\xi|
H|[l_i];v;\xi\rangle= \langle[l_i];v|aY_0 (\xi) + Y_+(\xi) +  Y_- +\tilde C
|[l_i];v\rangle$.  At same time an appropriate quasiclassical dynamics, which
is isomorphic to the exact quantum one when $sl_{pd}(2)=sl(2)$ [14-16], is
described by the classical Hamiltonian equations [5,14,16]
$$ \dot q =\frac{\partial {\cal H}}{\partial p}, \qquad \dot p =
-\frac{\partial {\cal H}}{\partial q}, \quad {\cal H}=
\langle z(t)|H|z(t)\rangle \eqno (3.2a)$$
for "motion" of the canonical parameters $p, q$ of the $SL(2)$ GCS
$|z(t)\rangle=\exp(-z(t) Y_++ z(t)^*Y_-)|\psi_0\rangle$ as trial functions
in the time-dependent Hartree-Fock variational scheme with the Lagrangian
${\cal L}=\langle z(t)|(i\partial/\partial t - H)|z(t)\rangle$; $p=
j\cos \theta, q= \phi, z = \theta/2 \exp (-i\phi)$ for $su(2)$ and
$p=j\cosh \theta, q= \phi, z = \theta/2 \exp (-i\phi)$ for $su(1,1)$.
An equivalent formulation in ${\bf Y}= (Y_1,Y_2,Y_0)$ space can be given
in terms of $sl(2)$ Euler-Lagrange equations,
$$ \dot {\bf y}=\frac{1}{2} {\bf \bigtriangledown} {\cal H}\times
{\bf \bigtriangledown} C, $$
$$ C=\pm y_0^2+y_1^2+y_2^2, \; y_i = \langle z(t)|Y_i|z(t)\rangle,
\; y_{\pm}=y_1\pm y_2, \; {\bf \bigtriangledown} =
(\partial/\partial y_1, \partial/\partial y_2, \partial/\partial y_0)
                 \eqno (3.2b)$$
reducing to the well-known (linear) Bloch equations [5,14,17].

Similarly, general ideas of the analysis above and calculation schemes
(3.1), (3.2) may be extended to the case of arbitrary polynomial algebras
$sl_{pd}(2)$ by using the energy functional ${\cal H}([l_i];v;\xi)=
\langle[l_i];v|\tilde{H}_{[l_i]}(\xi)|[l_i];v\rangle$ with
$\tilde{H}_{[l_i]}(\xi)$ being given by Eq. (2.9). Naturally, results
obtained in such a manner are not expected to coincide with exact solutions
on all subspaces $L([l_i])$ due to an essential nonlinearity of Hamiltonians
(2.4) and their non-equivalence  (unlike Eq. (2.7b)) to  diagonal parts of
Eq. (2.9); however, they yield, evidently, most close to exact "smooth"
(analytical) solutions (cf. [5,14]). Without dwelling on a discussion of all
aspects of such an extension we consider in detail an application of the
procedure (3.1) to the most wide-spread class [11] of Hamiltonians (2.4)
with the $su(2)$ dynamic symmetry which includes the model (2.3).

Note that the condition (3.1a) gives $e^{i\theta}=g/|g|$ as in the
linear case, and, due to the form of trial functions, it is sufficiently to
solve Eq. (3.1b) only for finding ground states $|[l_i]; v=0;\xi\rangle$.
Then, expanding r.h.s. of Eq. (2.5) in $Y_{\alpha}$ power series  and
taking into account defining relations for the $su(2)$ algebra one gets
after some algebra the following expressions
$$ E^{su(2)}_v ([l_i];\xi_0) ={\cal H}([l_i];v;\xi_0)
=C+a(l_0+j)+a(-j+v)\cos 2r-2|g|\sum_{f\geq 0} E_f^{\phi}(r;j;v), $$
$$E^{\phi}_f(r;j;v)=E^{\phi}_f(r;j;0)(\frac{1}{2}\sin 2r)^{-2v}\frac{(f)!
(2j-v)!(f+1)!}{(2j)!v!(f-v)!(f+1-v)!}\times$$
$$F(-v,-v+2j+1;f-v+1; \sin^2 r)F(-v,-v+2j+1;f-v+2; \sin^2 r),$$
$$E^{\phi}_f(r;j;0)=(\cos^{4j} r)\frac{(\tan r)^{2f+1}(2j)!}{(f)!(2j-f-1)!}
\sqrt{\phi_{n-2}(-j+f)}, \quad \phi_{n-2}(-j+f)=\frac{\psi(l_0+1+f)}
{(2j-f)(f+1)},   \eqno (3.3)$$
with $F(...)$ being the Gauss hypergeometric function [18], for energy
eigenvalues $E^{su(2)}_v ([l_i];\xi_0=rg/|g|)$ where diagonalizing values of
the parameter $r$ are determined from solving the algebraic equation
$$0 = \sum_{f\geq 0}\frac{\alpha^{2f}}{(2j-1-f)!f!}\{\frac{a \alpha}
{|g|}-[4\alpha^2 j -(1+\alpha^2)(2f+1)]\sqrt{\phi_{n-2}(-j+f)}\},
\quad \alpha =-\tan r \eqno (3.4)$$
For the case of the $su(1,1)$ dynamic symmetry Eqs. (3.3), (3.4), retaining
their general structure form, are slightly modified due to differences in
the definition (2.5) of $S_Y(\xi)$ for $su(2)$ and $su(1,1)$.
Let us make some remarks concerning this result.

1) As is seen from Eq. (3.3), its general structure coincides with the energy
formula given by the algebraic Bethe ansatz [6], and spectral functions
$E^{\phi}_f(r;j;v)$ are non-linear in the discrete variable $v$
labeling energy levels that provides a non-equdistant character of energy
spectra within  fixed subspaces $L([l_i])$ at $d([l_i])>3$. Besides,
due to the square roots in expressions for these functions different
eigenfrequencies $\omega^{su(2)}_v\equiv E^{su(2)}_v/\hbar$ are
incommensurable: $\; m\omega^{su(2)}_{v_1}\neq n\omega^{su(2)}_{v_2}$ that
is an indicator of an origin of collapses and revivals of the Rabi
oscillations [2,8] as well as of pre-chaotic dynamics [19]. Note that this
dependence is impossible to get by using GCS related to uncoupled subsystems.

2) The r.h.s. of Eq. (3.4) is a
polynomial of the degree $2j+1=d([l_i])$, and, in general, Eq. (3.4) may have
$2j+1$ different roots $r_i$ corresponding to $2j+1$ different stationary
values of the energy functional ${\cal H}([l_i];v;\xi)$. Therefore, one may
assume that it is possible to get more simple expressions for
$E^{\phi}_f(r;j;v)$ with any $v$ using $E^{\phi}_f(r;j;0)$ with different
roots $r_i$. Note that this conjecture is valid for little dimensions
$d([l_i])$ when Eqs. (3.3)-(3.4) give exact results. Another way to modify
and to simplify the results above is in using different properties,
including integral representations, of the hypergeometric functions
$F(a,b;c;x)$; specifically, using relations between hypergeometric
functions [18], one can express spectral functions $E^{\phi}_f(r;j;v)$ in
terms of the hypergeometric functions ${_4F_3(...;1)}$ (which are
proportional to the $sl(2)$ Racah coefficients).

3) Evidently, Eq. (3.3) generalizes Eq. (2.8a) for the (quasi)equidistant
approximation abovementioned. Indeed, when replacing functions
$\phi_{n-2}(-j+f)$ by their certain "average" values, series in (3.3), (3.4)
are summed up, and Eq. (3.3) is reduced to Eq. (2.8a); Taylor series
expansions of functions $\sqrt{\phi_{n-2}(-j+f)}$ provide  perturbative
corrections related to higher degrees of the an-harmonicity of Hamiltonians
(2.4). Furthermore, we can get an intermediate  approximation for energy
spectra if replacing in Eqs. (3.1) the exact energy functionals
${\cal H}([l_i];v;\xi)=\langle[l_i];v|\tilde{H}_{[l_i]}(\xi)|[l_i];v\rangle$
by their mean-field (corresponding to the Ehrenfest theorem) approximations
$${\cal H} ^{mfa}([l_i];v;\xi) =
a<Y_0(\xi)> + <Y_+(\xi)>\tilde g(<Y_0(\xi)>) + \tilde g^+(<Y_0(\xi)>)
<Y_-(\xi)> +\tilde C,$$
$$ <Y_{\alpha}(\xi)> =\langle[l_i];v|Y_{\alpha}(\xi)|[l_i];v\rangle
   \eqno (3.5) $$
Then Eqs. (3.3)-(3.4) are very simplified retaining their main
characteristic features. For example, for the model (2.3)  we  find
$$ E_v^{mfa} ([l_i];\xi_0) =$$
$$C+a(l_0+j)+a(-j+v)\cos 2r-
2|g|(j-v)\sin 2r \sqrt{(-j+v)\cos 2r +j+|k|+1}  \eqno (3.6a)$$
where $r$ is determined from the equation
$$\frac{a}{2|g|}\sin 2r= \cos 2r \sqrt{2j\sin^2 r +|k|+1} + \frac{j\sin^2 2r}
{2\sqrt{2j\sin^2 r +|k|+1}}    \eqno (3.6b)$$
(Similar expressions can be found for the point-like Dicke and the second
harmonic generation models.) Besides, substituting Eq. (3.5) in Eqs. (3.2)
one may get a mean-field approximation for dynamics equations reducing in the
${\bf Y}$ space representation to non-linear Bloch equations (cf.[5,11])
obtained from Eqs. (3.2b) by the substitution
$${\bf \bigtriangledown} {\cal H}= ([g+g^*] [\phi_{1}(y_0)]^{1/2},
[g-g^*] [\phi_{1}(y_0)]^{1/2}, a+ \frac{1}{2}[g(y_1+y_2)+g^*(y_1-y_2)]
[\phi_{1}(y_0)]^{-1/2}), $$
$${\bf \bigtriangledown} C = 2(y_1, y_2, y_0),
\quad \phi_{1}(y_0)= y_0+j+|k|+1    \eqno (3.7)$$

4) Finally, Eqs. (3.3) and (3.6a) can be used for obtaining appropriate
approximations 
$$U^{su(2)/mfa}_{H}(t)= \sum _{[l_i], v} S_Y(\xi_0)^{\dagger}
\; \exp (\frac{-it \omega^{su(2)/mfa}_v}{\hbar}) \;|[l_i];v\rangle
\langle[l_i];v| \;S_Y(\xi_0)
    \eqno (3.8)$$
for the evolution operators which are transformed to the form (1.4a) with
the help of the standard group-theoretical technique [20].

\section{Conclusion}

So, we have obtained new approximations for energy spectra and evolution
equations of models (1.1) by means of using the mapping (2.1) and the
variational schemes (3.1), (3.2) with
the $SL(2)$ GCS as trial functions. They may be called as a "smooth" $sl(2)$
quasiclassical approximations since they, in fact, correspond
to picking out "smooth" (analytical) $sl(2)$ factors $\exp(\xi_0 Y_+-
\xi_0^* Y_-)$ in exact diagonalizing operators $S(\tilde{\xi})$ and in
the evolution operator $U_H(t)$. These
approximations may be used for calculations of evolution of different
quantum statistical quantities (cf. [8,14]) and for determining bifurcation
sets of non-linear Hamiltonian flows in parameter space (cf. [5]).

Further investigations may be related to a search of
suitable multi-parametric specifications of exact diagonalizing operators
$S(\tilde{\xi})=S([\xi_0,\xi_1,\xi_2,...])$ using $\exp(\xi_0 Y_+-
\xi_0^* Y_-)$ as initial ones in iterative schemes which are
similiar to those developed to
examine non-linear problems of classical mechanics and optics [21] or
as "principal" factors  in the  diagonalization schemes like (2.7) for
Hamiltonians (1.1). From the practical point of view an important question
is  to get estimations of accuracy of approximations obtained and to make
comparisons of their efficiency with other approximations (e.g., given
in [8,9,11]). For the model (2.3) (and other ones with the
structure polynomial $\psi_3(x)$ of the third degree) it is of interest
to compare results of approximations found above with exact calculations
obtained by considering solvable cases of models under study. One of latters
is given by integral solutions [12] and other may be yielded by the
Riccati equations arising from a differential realization of $sl_{pd}(2)$
generators $V_{\alpha}$ [13]:
$$ V_-=d/dz,\; V_0=zd/dz+l_0,\; V_+=\psi_n(zd/dz+l_0)(d/dz)^{-1}
\eqno (4.1)$$
which is, in turn, related to a realization of $sl_{pd}(2)$ generators
$V_{\alpha}$ by quadratic forms in $sl(2)$ generators $Y_{\alpha}$
(cf. [15,22]). (In fact, this realization was used implicitly for
obtaining exact integral solutioms [12].) Besides, it is also of interest
to investigate possible connections of these results with quasi-exactly
solvable $sl(2)$ models [23,24].

The work along these lines is now in progress.

\section{Acknowledgements} 

Preliminary results of the work were reported at the VII International
Conference on Symmetry in Physics (JINR, Dubna, July 10-16, 1995) and at
the XV Workshop on Geometric Methods in Physics (Bialowieza, Poland, July
1-7, 1996). The author thanks C. Daskaloyannis, S.M. Chumakov and A.
Odzijewicz for useful discussions. The paper is prepared under partial
support of the Russian Foundation for Basic research, grant No 96-02 18746-a.

\end{document}